# Role of structure in the α and β dynamics of a simple glass-forming liquid

D. Fragiadakis and C. M. Roland

*Naval Research Laboratory, Chemistry Division, Washington, DC 20375-5342*

(December 12, 2016)

The elusive connection between dynamics and local structure in supercooled liquids is an important piece of the puzzle in the unsolved problem of the glass transition. The Johari-Goldstein β relaxation, ubiquitous in glass-forming liquids, exhibits mean properties that are strongly correlated to the long-time α dynamics. However, the former comprises simpler, more localized motion, and thus has perhaps a more straightforward connection to structure. Molecular dynamics simulations were carried out on a two-dimensional, rigid diatomic molecule (the simplest structure exhibiting a distinct β process) to assess the role of the local liquid structure on both the Johari-Goldstein β and the α relaxation. Although the average properties for these two relaxations are correlated, there is no connection between the β and α properties of a given (single) molecule. The propensity for motion at long times is independent of the rate or strength of a molecule's β relaxation. The mobility of a molecule averaged over many initial energies, a measure of the influence of structure, was found to be heterogeneous, with clustering at both the β and α timescales. This heterogeneity is less extended spatially for the β than for the α dynamics, as expected; however, the local structure is the more dominant control parameter for the β process. In the glassy state, the arrangement of neighboring molecules determines entirely the relaxation properties, with no discernible effect from the particle momenta.

## INTRODUCTION

The origin of the remarkable change in behavior as a liquid vitrifies is still an unsolved problem in condensed matter physics. The dynamics in the supercooled regime is spatially heterogeneous, and given the absence of structural heterogeneity, ferreting out the origin of this dynamic heterogeneity would seem to be paramount to solving the glass transition problem. A basic issue is the effect of the local liquid structure on the dynamics [1,2,3,4]. Implicit in drawing analogies from colloidal systems is the assumption that the arrangement of neighboring molecules largely determines the relaxation properties [5,6]. Molecular dynamics simulations, which enable the structure and dynamics to be followed in space and time with arbitrary precision, are well suited to test this assumption.

Isoconfigurational ensembles [7,8] have been an especially useful simulation methodology to assess structure-dynamics correlations. Multiple trajectories are simulated having the same initial particle positions but initial velocities randomly chosen from a Boltzmann distribution. At long times, particle mobilities, averaged over these trajectories (the so-called dynamic propensity) still show substantial heterogeneity; thus, the instantaneous velocities do not govern the mobilities. The heterogeneity must arise at least in part from "something" in the structure that predisposes some particles to have higher mobility than others [9]. What is this "something"? The isoconfigurational ensemble methodology is order agnostic; it cannot identify any specific structural feature responsible for controlling dynamics. In certain systems [10,11,12] regions of slow dynamics were found to be correlated with ordered clusters of particles [13,14], certain local geometrical motifs [15,16,17], the local composition [18], or soft localized modes of the quenched configuration [19]. The picture that emerges is that while there is a correlation between structure and dynamics, stronger at low temperature [20], the strength of such correlations is highly system dependent [21]. Quantities such as local density [22] or potential energy [23,24] do not in general correlate well with the spatial distribution of the dynamics. Two particularly interesting approaches have been the use of concepts from information theory to quantify links between structural and dynamical properties [25,26], as well as machine learning algorithms to identify new such links [27,28,29].

Work to date has focused on the relation of structure to dynamics at the timescale of the alpha relaxation and in the short-time caging regime [23,30]. In experimental studies of glass-forming materials, an additional relaxation, the Johari-Goldstein (JG) β process, is almost universally present at intermediate timescales. JG motion involves all atoms in the molecule and is observed even in systems with a completely rigid molecular structure. There is a large amount of evidence supporting a close relationship between this β and the α dynamics, and it has been suggested that the β process is simpler, more fundamental, non-cooperative (or weakly cooperative) motion that evolves over time into the cooperative α relaxation [4]. If this is the case, we might expect a more straightforward connection of the β relaxation to local structure, which could clarify any relationship between the long-time dynamics and the structure.

In this work we use molecular dynamics simulations to investigate the connection of the β relaxation to local structure. Model systems most often used to study glassy dynamics are mixtures of spherical particles, which do not show a JG relaxation, at least up to the timescales accessible by molecular dynamics simulations. Instead, we simulate a rigid, asymmetric dumbbell-shaped molecule, the simplest shape that captures the characteristics observed experimentally for the JG process [31,32,33,34]. The simulations were done for a two dimensional molecule, so that clustering can be more clearly depicted. We find that this 2D behavior is qualitatively the same as for the three dimensional case.

## METHODS

We studied in two dimensions a binary mixture of N=4000 (2600:1400) rigid dumbbell-shaped molecules labeled AB and CD. Atoms belonging to different molecules interact through the Lennard-Jones potential

$$U_{ij}(r) = 4\epsilon_{ij}\left[\left(\frac{\sigma_{ij}}{r}\right)^{12} - \left(\frac{\sigma_{ij}}{r}\right)^{6}\right] \quad (1)$$

where $r$ is the distance between particles, and $i$ and $j$ refer to the particle types A, B, C and D. The energy and size parameters are based on the Kob-Andersen (KA) binary mixture of spheres, which does not easily crystallize [35]. This was done in the same way as in the 3D systems in refs. [31,32,33,34]; to wit, the energy parameters $\epsilon_{ij}$ are those of the KA liquid, i.e., $\epsilon_{AA} = \epsilon_{AB} = \epsilon_{BB} = 1.0$, $\epsilon_{CC} = \epsilon_{CD} = \epsilon_{DD} = 1.0$, and $\epsilon_{AC} = \epsilon_{AD} = \epsilon_{BC} = \epsilon_{BD} = 1.5$. To set $\sigma_{ij}$, we use the original KA parameters for the larger A and C particles, while the smaller B and D particles have a size 62.5% that of A and C, respectively. Thus, $\sigma_{AA} = 1$, $\sigma_{CC} = 0.88$, $\sigma_{BB} = 0.625$, and $\sigma_{DD} = 0.625 \times 0.88$. For the interactions between different types of particles, we take $\sigma_{ij} = K_{ij}(\sigma_{ii} + \sigma_{ij})$ where $K_{ij}$=0.5 (additive interaction) when the particles belong to the same type of molecule (i, j = AB, CD) and $K_{ij} = 0.4255$ when the particles belong to different molecule types, the latter chosen to give the KA value for $\sigma_{AC} = 0.8$. All atoms have a mass $m$ = 1. The bond lengths A-B and C-D were fixed to d=$\sigma_{AA}$/3. In two dimensions we use a 65:35 ratio of AB to CD molecules instead of the usual 80:20 ratio used in 3D; we found this to be necessary to suppress crystallization and phase separation, as is the case for the KA mixture of spheres [36]. Unless otherwise noted, we follow the dynamics of the AB species. The data are presented in normalized units (Lennard-Jones units) of length $\sigma_{AA}$, temperature $\epsilon_{AA}/k_B$ and time $(m\sigma_{AA}^2/\epsilon_{AA})^{1/2}$.

Simulations were carried out using GROMACS, with the velocity Verlet algorithm, a Nose-Hoover thermostat, and Parrinello-Rahman barostat [37,38,39] at a pressure of P=1, in a square box with periodic boundary conditions. Bond lengths were maintained constant using the LINCS algorithm [40]. At each temperature the system was equilibrated for the shorter of $10\tau_\alpha$ or $t_{eq} = 5 \times 10^6$. At low temperatures, T<0.3, the $\alpha$ relaxation time is much longer than the total (equilibration and production) simulation time, whereby the system is out of equilibrium (i.e., a glass). In the glassy state, translational and orientational correlation functions do not decay to zero over the duration of the simulation runs. For the simulations at three temperatures in the glassy state, production run times were much less than $t_{eq}$, ensuring an absence of significant aging (drift in pressure, potential energy, or dynamical correlation functions) during a run.

At the glass transition defined by incomplete decay of the correlation functions, the α relaxation time is about eight orders of magnitude longer than the vibrational relaxation times. For a real glass-forming liquid this corresponds to time scales in the range of $10^{-4}$ s, rather than the more usual $\tau_\alpha \sim 100$ s for experimental glass transitions.

The supercooled dynamics of mixtures of two-dimensional Lennard-Jones particles differ from their three-dimensional counterpart in the extent of transient localization of particles and the degree of decoupling of translational and orientational correlations [41]. In the system studied herein, the temperature dependence of the relaxation times, strengths, and spectral shapes for both the α and β processes, as well as the nature of their dynamic heterogeneity, are qualitatively identical to those for the three-dimensional asymmetric dumbbell system (representative comparison shown below).

## RESULTS AND DISCUSSION

We follow reorientational motions via the first order rotational correlation function

$$C_1(t) = \langle \cos\theta(t) \rangle \qquad (2)$$

where $\theta$ is the angular change of a unit vector along the molecular axis. The same information is contained in the corresponding susceptibility $\chi(\omega)$, which can be more readily compared to experimental dielectric spectra. The susceptibility is calculated from

$$\begin{aligned}\chi(\omega) &= \chi'(\omega) + i\chi''(\omega) \\ &= 1 + i\omega \int_0^\infty dt\, e^{i\omega t} C_1(t)\end{aligned} \qquad (3)$$

Figure 1 shows $C_1(t)$ and $\chi(\omega)$ for various temperatures. There is a small initial drop in $C_1$ at $t \sim 0.1$, corresponding to rattling of the molecules within the local cage formed by neighboring molecules. At high temperatures, $C_1$ then decays to zero via a single step. Below a temperature $T_{on}$, the relaxation occurs in two steps, the shorter-time β and longer-time α. The latter grows in intensity with decreasing temperature at the expense of the β relaxation strength. At or below $T \sim 0.3$ ($\equiv T_g$), the α relaxation time is much longer than the total simulation run time, and the system is in the non-equilibrium glassy state. The β process remains as a broad step, with $C_1$ no longer decaying to zero. The susceptibility $\chi(\omega)$ contains the same information as $C_1(t)$: a small high-frequency peak (at low temperatures), or change in slope (at high temperatures) corresponding to caged dynamics; a peak corresponding to the β process; and below $T_{on}$ the α dispersion developing as a weak shoulder that intensifies on cooling. Higher-order rotational correlation functions, as well as translational relaxation, are not shown but behave in a qualitatively similar way to $C_1(t)$, with slightly different relative intensities and relaxation times for the α, β, and vibrational dynamics [32].

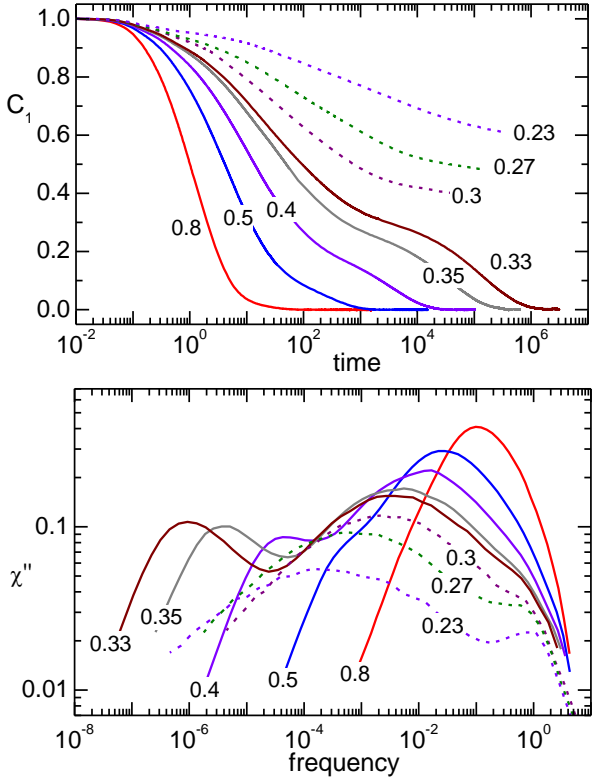

FIG. 1. (top) First order rotational correlation function as a function of time; (bottom) imaginary part of the first-order rotational susceptibility as a function of frequency. The temperatures are indicated in units of $\epsilon_{AA}/k_B$. Dashed lines for $T \leq 0.3$ denote the glassy state.

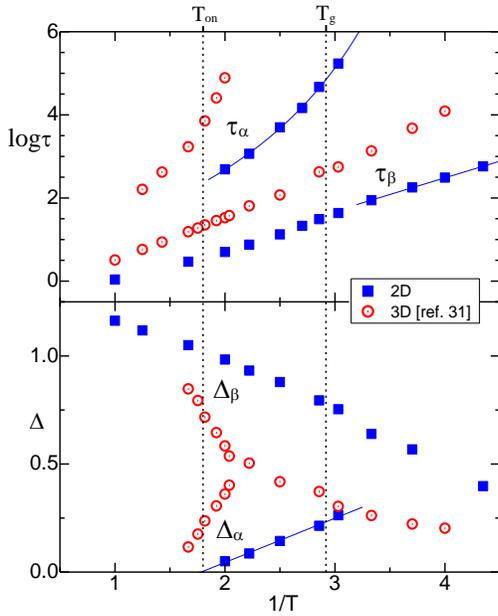

FIG. 2. Relaxation times (top) and strengths (bottom) of the α and β processes, obtained by fitting the data of Figure 1 (squares). Lines through $\tau_\alpha$, $\tau_\beta$ and $\Delta_\alpha$ are respectively fits of eq. 5, the Arrhenius equation, and linear behavior. The onset and glass transition temperatures are indicated by vertical dashed lines. Also shown are results (circles) from simulations of a similar three-dimensional molecule [32].

The contributions to the relaxation spectrum from $C_\alpha(t)$ and $C_\beta(t)$ of the respective α and β processes were deconvoluted using the Williams ansatz [42]:

$$C_1(t) = \Delta_\alpha C_\alpha(t) + \Delta_\beta C_\alpha(t) C_\beta(t) \quad (4)$$

where Δ represents the relaxation strength. We used a stretched exponential function [43] for the $\alpha$ relaxation and a Cole-Cole function [43], or its Fourier transform to the time-domain, for the β relaxation. Eq. 4 assumes the β dynamics takes place in an environment that is rearranging on the time scale of the $\alpha$ process, with the two being "statistically independent". A common alternative is to assume additivity of the two relaxations in the time or frequency domain. The β relaxation must then be described with an asymmetric function to provide a satisfactory fit. This requires an additional adjustable parameter, yet still gives poorer fits and larger uncertainties than the Williams ansatz.

Figure 2 shows the variation with temperature of the relaxation times and strengths. The α relaxation

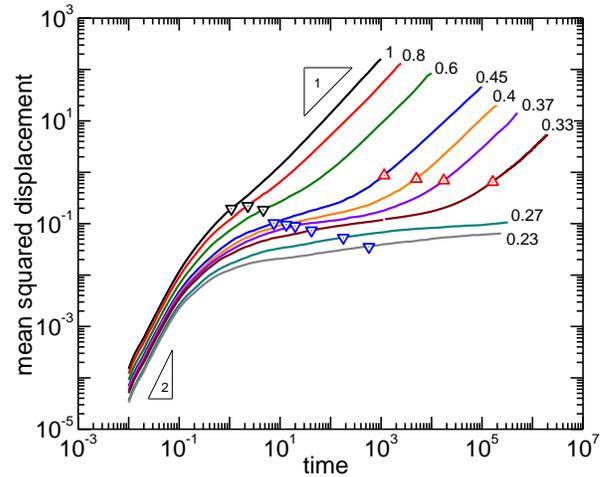

FIG. 3. Mean square displacement of the molecular center of mass at the indicated temperatures. Symbols indicate the α (triangles) and β (inverted triangles) relaxation times.

time can be described by the Vogel-Fulcher-Tammann equation [43]

$$\tau_\alpha(T) = \tau_0 \exp\left(\frac{B}{T-T_0}\right) \quad (5)$$

where $\tau_0$, $B$ and $T_0$ are constants. The β relaxation time shows Arrhenius behavior ($T_0=0$) in the glass, which changes to a slightly stronger, non-Arrhenius temperature dependence above $T_g$. The β relaxation strength increases with increasing temperature, while $\Delta_\alpha$ decreases, becoming zero at $T_{on} = 0.56$. (That is, at high temperatures the structural dynamics are relatively unconstrained.) Included in Fig. 2 are the corresponding relaxation times for the 3 dimensional case [32], showing qualitatively the same behavior.

In Figure 3, $\tau_\alpha$ and $\tau_\beta$ are indicated on plots of the mean square displacement of the center of mass. The behavior is typical for glass-forming liquids, with an initial slope of 2 (ballistic motion) and a long-time slope of unity indicating diffusive behavior. Below the α onset temperature, the α relaxation time corresponds to an approximately constant value of the mean square displacement, in the beginning of the diffusive regime; this is a consequence of the coupling of the α reorientations to diffusion. At lower temperatures a plateau is evident; that is, a broad shallow step centered around $\tau_\beta$. This plateau reflects

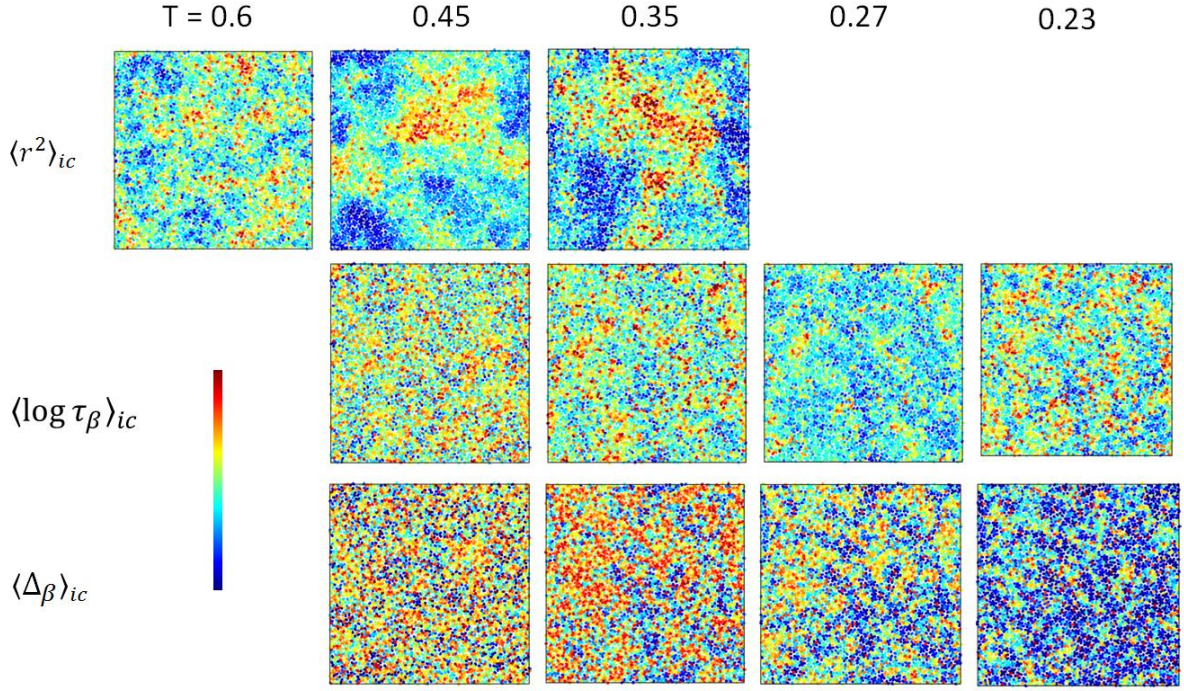

FIG. 4. Spatial distribution of (top) the dynamic propensity at $t = 1.5\tau_\alpha$, (middle) the isoconfigurationally averaged $\beta$ relaxation time, and (bottom) relaxation strength. The color scale is normalized to span the minimum to the maximum values for each panel.

small displacements caused by the rotational jumps comprising the β process [31].

From the simulations in the isoconfigurational ensemble we determined for each molecule the dynamic propensity $\langle r^2 \rangle_{ic}$. This is defined as the squared displacement of the center of mass at a reference time $t_{ref} = 1.5\tau_\alpha$, averaged over 100 runs, each starting from the same particle configuration with randomized initial velocities [7,8]. Figure 4 (top) shows a map of the dynamic propensity for three temperatures in the liquid state. Similar to previous studies [7,8,23], there is substantial heterogeneity, with molecules having large and small $\langle r^2 \rangle_{ic}$ organized into clusters, the size of which increases on cooling [44].

As calculated herein, $\langle r^2 \rangle_{ic}$ characterizes the propensity of a molecule for motion at long times, $t > \tau_\alpha$. To investigate the dynamics on the β timescale, we calculated from the first-order rotational correlation function for each molecule an isoconfigurationally averaged logarithm of the β relaxation time $\langle \log \tau_\beta \rangle_{ic}$ and β relaxation strength $\langle \Delta_\beta \rangle_{ic}$. (Note that because the β relaxation manifests as only a very weak step in the mean square displacement, using the latter to compute the dynamic propensity is impractical). A correlation function $\langle C_1(t) \rangle_{ic}$ was first calculated for each molecule, by averaging its rotational correlation function over a suitable time interval $(0, t_{ave})$ across the isoconfigurational ensemble. The time $t_{ave}$ is intermediate between $\tau_\alpha$ and $\tau_\beta$; that is, long enough that the entirety of the β relaxation is captured in the calculated correlation function, but sufficiently shorter than $\tau_\alpha$ so that the structure (which we are trying to correlate to the β dynamics) does not change significantly. As long as these conditions hold, the exact value of $t_{ave}$ does not significantly affect the results of our calculation.

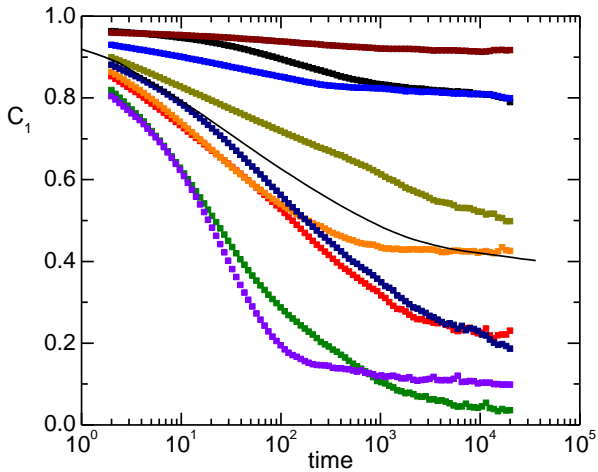

FIG. 5. Representative isoconfigurationally averaged correlation functions $\langle C_1(t) \rangle_{ic}$ for ten individual molecules at T=0.3 (symbols). The solid line is the mean correlation function for all molecules.

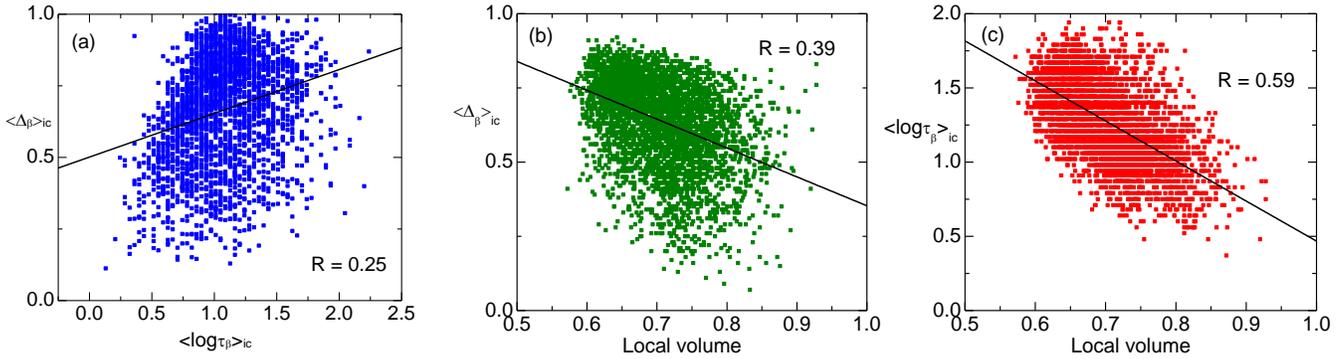

FIG. 6. (a) Correlation of isoconfigurationally averaged β relaxation strength and relaxation time, along with the correlation of the local volume with the isoconfigurationally averaged (b) β relaxation strength and (c) β relaxation time (T= 0.3)

Representative isoconfigurationally averaged single-molecule correlation functions, along with the total (averaged over all molecules) $C_1(t)$, are shown in Figure 5. If structure had no influence on the β dynamics, the curves for individual molecules would collapse onto $C_1(t)$. They do not; instead, a step decrease corresponding to the β relaxation is observed at widely different timescales and with a range of relaxation strengths.

A suitable function (the time-domain transform of the Cole-Cole equation) was fit to the $\langle C_1(t) \rangle_{ic}$ for each molecule to extract an isoconfigurational relaxation time $\langle \log \tau_\beta \rangle_{ic}$ and relaxation strength $\langle \Delta_\beta \rangle_{ic}$. These quantities are depicted in Figure 4. While some clustering of molecules with similar $\langle \log \tau_\beta \rangle_{ic}$ and similar $\langle \Delta_\beta \rangle_{ic}$ is present, these clusters are much smaller than those for $\langle r^2 \rangle_{ic}$; moreover, the cluster size does not grow appreciably on cooling. In fact there does not seem to be any similarity between the propensity maps at the α and β relaxation timescales; that is, a molecule's propensity for motion at long times is independent of its propensity for having a fast or strong β relaxation. This result calls to mind our previous finding that in a single simulation run (i.e., without the isoconfigurational averaging), single-molecule β relaxation times and strengths are independent of each other, and independent of the single-molecule α relaxation time [33]. This behavior at the single-molecule level is in striking contrast to the many correlations between average properties of the α and β processes, which has led to the inference of a direct relationship between the two [45,46]. $\langle \log \tau_\beta \rangle_{ic}$ and $\langle \Delta_\beta \rangle_{ic}$ are also mutually uncorrelated (Figure 6a), as found for the non-averaged values [33].

To quantify the clustering, we calculate a length scale for the dynamic propensity as follows. The standard deviation $\sigma(0)$ of the propensity is a measure of its heterogeneity. We then average ("blur") the propensity over a length scale L by replacing each molecule's propensity by the average value within a radius L of the molecule; this gives a new standard deviation $\sigma(L)$, which is smaller for larger values of L. We then define the characteristic length scale $\lambda$ as the value of $L$ for which the $\sigma(L)/\sigma(0)$ reaches a given value, here 1/e. Figure 7 shows the length scales extracted in this manner using the data in Figure 4. The length $\lambda_\alpha$, characterizing heterogeneity of the dynamic propensity for the α relaxation, is large (many times the molecular diameter) and increases strongly on cooling. Though uncorrelated with each other, both the β relaxation time and strength are characterized by a common length scale $\lambda_\beta$, equal to a couple molecular diameters and thus much smaller than $\lambda_\alpha$. $\lambda_\beta$ increases weakly with decreasing temperature, reaching a peak slightly below $T_g$ and then decreasing on further cooling. The decrease below $T_g$ may be related to the decrease in the 4-point correlation function at the β timescale [33].

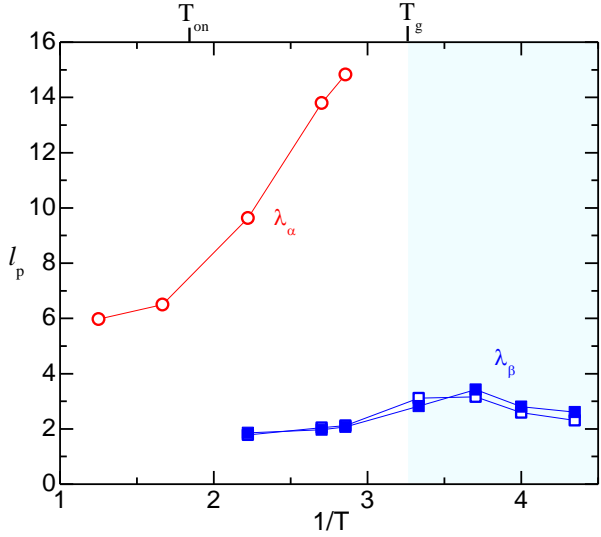

FIG. 7. Characteristic length scale of the dynamic propensity (circles), and the isoconfigurationally averaged β relaxation time (filled squares) and β relaxation strength (empty squares). The shaded area corresponds to the glassy state.

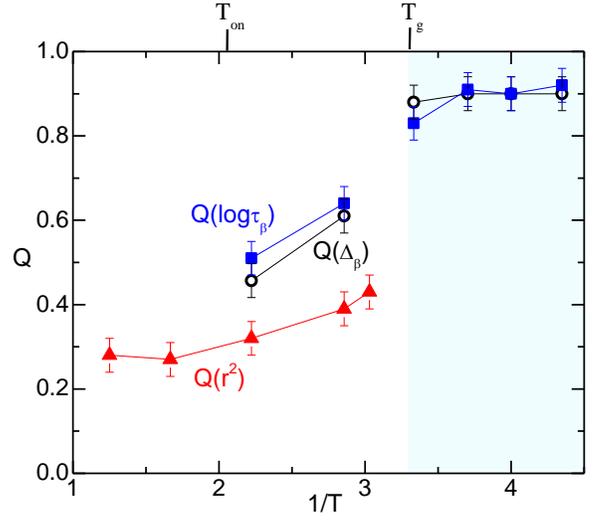

FIG. 8. Degree to which heterogeneity is determined by the structure, as calculated using eq. 8, for the dynamic propensity (triangles), β relaxation time (squares), and β relaxation strength (circles). The shaded area corresponds to the glassy state.

When discussing the origin of the JG relaxation, regions of looser packing, so-called "islands of mobility" [47], are sometimes invoked. We partition the simulation box into Voronoi polyhedra [48] centered on the molecular centers of mass, and use the volume of the polyhedron as a measure of the volume local to each molecule. Only a very weak (R=0.39) correlation is present between $\langle \Delta_\beta \rangle_{ic}$ and this local volume (Fig. 6b), but surprisingly this suggests that molecules with smaller local volume are slightly more likely to have a more intense β relaxation. A somewhat better but still weak correlation (R=0.59) exists between local volume and the β relaxation time (Fig. 6c); to wit, looser packing is associated with faster β dynamics.

Nevertheless, local structure clearly plays a determinative role in the β relaxation. We quantify the extent to which structure (as opposed to initial momenta) controls dynamics as follows: By analyzing only a single simulation run, we obtain N relaxation times $\log \tau_\beta^{(i)}$ of individual molecules (*i* is the molecule index), which form a distribution. The breadth of this distribution, or amount of heterogeneity, can be quantified by the standard deviation of $\log \tau_\beta^{(i)}$, i.e.,

$$S_1 = \left[ N^{-1} \sum_i \left( \log \tau_\beta^{(i)} - \log \tau_\beta \right)^2 \right]^{1/2}. \quad (6)$$

A portion of the heterogeneity will be due to the variation in local structure around each molecule in the starting configuration, the rest due to the variations in initial velocities. When we average over multiple simulation runs, each with identical starting configurations but different initial velocities, we obtain N isoconfigurationally averaged single molecule relaxation times $\langle \log \tau_\beta^{(i)} \rangle_{ic}$. These form a different distribution, the breadth of which is only due to variation in local structure. The standard deviation,

$$S_{iso} = \left[ N^{-1} \sum_i \left( \langle \log \tau_\beta^{(i)} \rangle_{ic} - \log \tau_\beta \right)^2 \right]^{1/2} \quad (7)$$

must therefore be smaller than $S_1$. Taking the ratio of the two standard deviations

$$Q(\log \tau_\beta) = \frac{S_{iso}}{S_1} = \left[ \frac{\sum_i \left( \langle \log \tau_\beta^{(i)} \rangle_{ic} - \log \tau_\beta \right)^2}{\sum_i \left( \log \tau_\beta^{(i)} - \log \tau_\beta \right)^2} \right]^{1/2} \quad (8)$$

gives us a quantity that reflects to what extent $\log \tau_\beta$ is determined only by local structure. The same analysis was performed for $\Delta_\beta$ to yield $Q(\Delta_\beta)$, and for the squared displacement at $t_{ref} = 1.5\tau_\alpha$, $Q(r^2)$ characterizing the dynamic propensity at the α relaxation timescale. These three ratios of standard deviations are plotted in Figure 8. At a given temperature, $Q(\log \tau_\beta) \sim Q(\Delta_\beta)$, and both are larger than $Q(r^2)$; i.e., the heterogeneity of the β relaxation is determined by the structure to a greater extent than is that of the α relaxation. All three quantities increase on cooling, and below the glass transition the standard deviations for the β process both reach a plateau of ~0.95. This means that in the glass, heterogeneity of the β process is almost entirely due to variation of the local structure.

## CONCLUSIONS

The fact that the dynamic propensity is heterogeneous means that "something" in the structure determines the α relaxation dynamics [9], though its exact nature remains elusive and perhaps system-dependent. In a rigid dumbbell-shaped molecule that exhibits a JG β relaxation, we show that "something else" in the structure also governs the β relaxation. The local structure gives molecular propensities for faster or slower, as well as stronger or weaker, β dynamics; however, these propensities are independent from each other, and from the dynamic propensity at the α timescale. Structure controls the β relaxation properties to a larger extent than it affects motion at longer times, and this disparity increases with cooling to eventually approach 100%; that is, in the glassy state the instantaneous momentum of a particle has a negligible influence on its dynamics.

## ACKNOWLDEGMENT

This work was supported by the Office of Naval Research.